\documentclass[12pt]{article}
 \newcommand{\be}{\begin{equation}}
\newcommand{\ee}{\end{equation}} \newcommand{\bea}{\begin{eqnarray}}
\newcommand{\eea}{\end{eqnarray}}

 \newcommand{\Ahat}{\hat{A}}
\newcommand{\Bhat}{\hat{B}} \newcommand{\rhohat}{\hat{\rho}}
\newcommand{\Lamhat}{\hat{\Lambda}} \newcommand{\ihat}{\hat{i}}
 
\newcommand{\that}{\hat{t}} \newcommand{\xhat}{\hat{x}}
\newcommand{\yhat}{\hat{y}} 
   \newcommand{\adot}{\dot{a}} \newcommand{\addot}{\ddot{a}}
\newcommand{\bdot}{\dot{b}} \newcommand{\bddot}{\ddot{b}}
 \newcommand{\ndot}{\dot{n}}

\newcommand{\bfm}{\mathbf}
\newcommand{\ytilde}{\tilde{y}}
\newcommand{\ttilde}{\tilde{t}}

\begin{document}

\begin{titlepage}
\begin{flushleft}
       \hfill                      {\tt hep-th/0007254}\\ \hfill
       HIP-2000-39/TH \\ \hfill            December 18, 2000\\
\end{flushleft}
\vspace*{3mm}
\begin{center}
{\Large {\bf Constraints on Brane and Bulk Ideal Fluid \\} {\bf in
Randall-Sundrum Cosmologies\\}} \vspace*{12mm} {\large Kari
Enqvist\footnote{E-mail: kari.enqvist@helsinki.fi}, Esko
Keski-Vakkuri\footnote{E-mail: esko.keski-vakkuri@hip.fi} and
Syksy R\"{a}s\"{a}nen\footnote{E-mail:
syksy.rasanen@helsinki.fi}\\}

\vspace{5mm}

{\em {}$^{1,2}$Helsinki Institute of Physics \\ P.O. Box 9,
FIN-00014 University of Helsinki, Finland }

\vspace{3mm}

and

\vspace{3mm}

{\em {}$^{1,3}$Department of Physics \\ P.O. Box 9, FIN-00014
University of Helsinki, Finland}

\vspace*{10mm}
\end{center}

\begin{abstract} \noindent

We investigate constraints for including bulk and brane matter in
the Randall-Sundrum model. In static configurations with two
zero thickness branes, we find that no realistic brane
matter is possible. We also consider
the possibility that the radion has stabilized by dissipating its
energy into the bulk in the form of some unspecified matter, and
find the Randall-Sundrum cosmological solutions in the presence of
bulk ideal fluid. We discover that the metric  is necessarily in a static
configuration. We also discover that there is only one allowed equation of
state for the bulk fluid, $p=\rho$, corresponding to the stiff ideal fluid.
We find the corresponding brane cosmologies and compare them with the
Friedmann-Robertson-Walker model.

\end{abstract}

\end{titlepage}

\baselineskip16pt

\section{Introduction}

Randall and Sundrum have recently suggested a novel solution to
the hierarchy problem involving extra dimensions
\cite{Randall:1999ee, Randall:1999vf}, which has attracted much
attention. The RS proposal assumes a five-dimensional spacetime
where the extra dimension is a $S^1/Z_2$ orbifold, with branes
located at the two orbifold fixed points. The branes are thus at
the spatial boundaries of the bulk spacetime. The brane tensions
together with a negative bulk vacuum energy provide the source for
the 5-d Einstein equation, which yields as the solution the metric
\begin{equation} \label{RS}
  ds^2= e^{-2k|y|}\ (-dt^2+\sum^3_{i=1}(dx^i)^2) + dy^2 \ .
\end{equation}

The exponential warp factor rescales the physical masses on the
second ``TeV brane'', giving rise to a large suppression factor
with respect to the Planck mass, and hence a possible solution to
the hierarchy problem. However, this is achieved at the expense of
fine tuning: the cosmological constants on the two branes must be
of equal magnitude but with opposite signs. Also, the size of the extra
dimension, or the radion, is a four dimensional modulus field
whose value needs to be fixed at the right scale in order for the
RS scenario to work.

The RS model has been refined in many ways since its inception,
but the question of radion stabilization remains unsolved.
According to commonly accepted wisdom, moving branes would give
rise to particle masses which change in time \cite{Csaki:1999}, so
that the radion should be very nearly stabilized by the beginning
of nucleosynthesis so as not to conflict with observation. One
possibility was proposed by Goldberger and Wise, who assumed the
existence of a massive bulk scalar field with self-interactions on
the branes \cite{Goldberger:1999uk}; see also \cite{Cline:2000,
DeWolfe:1999}. By integrating over the fifth dimension one then
generates an effective potential for the radion, which has
a non-trivial minimum. As the radion field settles into the minimum,
the size of the extra dimension gets fixed.

Using a bulk scalar field to stabilize the radion has also been
studied in \cite{Kanti:2000a}. Other ideas for fixing the size of
the fifth dimension include gaugino condensation in a
supersymmetric setting \cite{Luty:1999} and the possibility that
the Hubble redshift might damp the radion so that it is almost
stabilized \cite{Steinhardt:1999}\footnote{In \cite{Kim:2000} the
interesting possibility that an asymptotically constant radion
emerges from the Einstein equation naturally without the
introduction of extra degrees of freedom was considered.
Unfortunately the Ansatz used actually precludes any dynamics for
the size of the fifth dimension.}. A phenomenological mechanism
for stabilizing the radion was considered in \cite{Csaki:1999}.

The implications of radion stabilization independent of the
specifics of the stabilization mechanism were considered in
\cite{Kanti:1999a, Kanti:1999b}. It was found that a fixed size
for the fifth dimension requires a certain non-constant form for
the $yy$-component of the bulk energy-momentum tensor. This form
turned out to be determined by a constraint that related the
$yy$-component of the stress tensor to its trace. This constraint
could be understood in terms of the backreaction of the radion to
the inclusion of matter.

Whatever the actual mechanism for stabilizing the radion, it is
natural to assume that the fixed size of the extra dimension
results from dynamical evolution. In the Goldberger-Wise scenario, the
radion would evolve from its initial state towards the minimum by
dissipating the extra energy away. The possibility of the radion
stabilizing via decay into Standard Model particles on the TeV
brane was considered in \cite{Csaki:1999}. However, it does not
seem farfetched to assume that there are other fields in the bulk,
which can couple to the radion, whether directly, via the GW field
or via some other mechanism, providing an alternative channel for
energy dissipation. Such additional fields could {\em e.g.} be a
part of the field content of a supergravity theory in the bulk.
The asymptotic final state in the bulk would then involve the
radion field with a constant value dictated by the minimum of the
effective potential, together with some bulk matter.

If bulk matter exists, at large times one would expect it to be in
the state of maximum entropy. This means that e.g. viscous flow
should eventually get damped away, and that the bulk matter can be
considered an ideal fluid. In this case, stress tensor of the bulk
fluid will be spatially isotropic, in particular the $yy$-component
will {\em not} differ from the other diagonal
spatial components, unlike in previous investigations
\cite{Kanti:2000a, Kim:2000, Kanti:1999a, Kanti:1999b, Kanti:2000b,
HBKim:2000, Mohapatra:2000, Kennedy:2000}.

In section 2 we find the general form of the metric assuming
homogeneity and isotropy with respect to the three visible spatial
dimensions, one static extra dimension and no matter flow along
this extra dimension. We then focus on two-brane systems with zero
thickness branes, and make a distinction between
static and non-static configurations. We investigate constraints on
brane matter in two-brane systems, and show that in static
configurations the only allowed forms of brane matter
are a cosmological constant and domain walls moving at the speed
of light. {\em Single}-brane configurations easily allow generic
matter, but we are interested in the possibility of adding another brane
necessary for generating the hierarchy. We then
comment on the scenarios for including realistic brane matter
in non-static {\em two}-brane configurations.
In section 3 we consider the case of an ideal
fluid plus a cosmological constant in the bulk and find the exact
cosmological solutions. Curiously, we find that the metric must be in
a static configuration, leaving no room for realistic brane matter,
and that there is only one allowed equation of state for the
bulk fluid, $p=\rho$, corresponding to the stiff ideal fluid.
We note that it is possible for both branes to have a positive
cosmological constant, and one brane may even have a zero
cosmological constant, a result discovered in \cite{Kaloper:1999}
and emphasized in \cite{Kanti:2000a}. In section 4 we discuss our results.

\section{Warped spacetimes bounded by two branes}

\subsection{General form of the metric}

The object of our interest is the Einstein equation in 4+1
dimensions,
\be
  G_{A B} = \kappa^2 T_{A B} \ .
\ee

\noindent In the above, $\kappa^2 = 1/M^3$, where $M$ is the
Planck scale in five dimensions. The indices $A$ and $B$ run
through time $t$, three spatial coordinates $x^i$ with infinite
range and a compact spatial coordinate $y$. We use the standard
notational convention and denote partial derivatives with respect
to $t$ and $y$ by dots and primes, respectively.

We begin with two general assumptions regarding the metric and the
matter content:

\begin{enumerate}

\item The spacetime is homogeneous and isotropic with respect to the three
spatial coordinates $x^i$.

\item There is no matter flow in the $y$-direction.

\end{enumerate}

\paragraph{Assumption 1.} The most general metric obeying
homogeneity and isotropy with respect to the spatial coordinates
$x^i$ is
\be \label{metric1}
  ds^2 = -n(t,y)^2 dt^2 + 2\, c(t,y)\, dt\, dy + b(t,y)^2 dy^2 +
\frac{a(t,y)^2}{\left(1+\frac{K}{4}r^2\right)^2} \sum^3_{i=1}(dx^i)^2 \ ,
\ee

\noindent where $r^2=\sum^3_{i=1}(x^i)^2$ and $6 K$ is the
constant three-dimensional spatial curvature. While the general form of
the metric includes a $dtdy$ cross term, most of the literature has adopted
Gaussian normal coordinates, where the cross term is absent ($c=0$). The same
assumption will be adopted in the present paper.
We call the metric with $c=0$ the \emph{cosmological} Randall-Sundrum metric in
contrast to the original static Randall-Sundrum solution. For simplicity of
notation, we put $K=0$ for the rest of this work. This does not affect our
essential results.

\paragraph{The Einstein Equation.}
The nontrivial components of the Einstein equation for the metric
(\ref{metric1}) in Gaussian normal coordinates, with $K=0$, are
\bea \label{einstein1}
  G_{tt} &=& 3\left[-\frac{n^2}{b^2}
\left( \frac{a''}{a}+
\frac{a'}{a}\left(\frac{a'}{a}-\frac{b'}{b}\right)\right) +
\frac{\adot}{a}\left(\frac{\adot}{a}+\frac{\bdot}{b}\right)\right]
= -\kappa^2 n^2 T^t_{\ t} \nonumber \\
  G_{ii} &=& \frac{a^2}{b^2} \left[2\frac{a''}{a}+\frac{n''}{n}+\frac{a'}{a}
\left(\frac{a'}{a}+2\frac{n'}{n}\right)-\frac{b'}{b}\left(\frac{n'}{n}
+2\frac{a'}{a} \right) \right] \nonumber \\
  &&-\frac{a^2}{n^2}\left[2\frac{\addot}{a}+\frac{\bddot}{b}+\frac{\adot}{a}
\left(\frac{\adot}{a}-2\frac{\ndot}{n}\right)
+\frac{\bdot}{b}\left(2\frac{\adot}{a}-\frac{\ndot}{n}\right)\right]
= \kappa^2 a^2  T^i_{\ i} \nonumber \\
  G_{yy} &=& 3\left[ \frac{a'}{a} \left( \frac{a'}{a} +\frac{n'}{n}\right)
-\frac{b^2}{n^2}\left(\frac{\addot}{a} + \frac{\adot}{a} \left(
\frac{\adot}{a} -\frac{\ndot}{n} \right)\right) \right] = \kappa^2
b^2 T^y_{\ y} \nonumber \\
  G_{ty} &=& 3\left[ \frac{n'}{n}\frac{\adot}{a}+\frac{a'}{a}\frac{\bdot}{b}
-\frac{\adot'}{a} \right] = -\kappa^2 n^2 T^t_{\ y}  \ .
\eea

We are interested in models with two parallel branes at the
endpoints of the $y$-coordinate. In particular, we demand that
the proper transverse distance between the branes is independent
of time. So we make one more assumption:

\begin{enumerate}

\item[3.] In the coordinates where the branes are at $y=y_1,y_2$, we have
$\bdot=0$.

\end{enumerate}

\paragraph{Assumptions 2 and 3.}  No matter flow in the $y$-direction
means that $T_{ty}=0$. Inserting this into the relevant component
of the Einstein equation (\ref{einstein1}), along with $\bdot=0$,
we obtain the result
\be \label{adoteqn}
  \adot(t,y) = A(t) n(t,y) \ ,
\ee

\noindent where $A$ is an unknown function.
We are interested in cosmological
solutions, and so will throughout the paper assume that $\adot\neq0$.

\subsection{Motivation for a static fifth dimension}

The assumption of a static fifth dimension appears in most papers
on the RS model. Since it does not seem obvious that a non-static fifth
dimension would in general lead to conflict with observation, we present a
brief discussion to motivate the assumption $\bdot=0$.

The RS model was first envisaged as a solution to the hierarchy
problem: the (four-dimensional) length scale increases as one
moves from one brane to the other, which induces a change in the
mass scales between the branes. The hierarchy between the Planck scale
on the (hidden) brane at $y=y_1$ and the TeV scale on the (observable)
brane at $y=y_2$ requires that the functions $n(t,y)$ and $a(t,y)$
satisfy
\bea \label{strict}
  \frac{n(t,y_1)}{n(t,y_2)} &=& N \nonumber \\
  \frac{a(t,y_1)}{a(t,y_2)} &=& N h(t) \ ,
\eea

\noindent where $N\sim 10^{16}$
is the ratio of the Planck and TeV scales and $h(t)$ is
some function of time (to allow for the possibility of different
cosmological expansion factors on the two branes). This is a
rather weak constraint, since it makes no reference to behaviour
away from the branes. Therefore, in the literature there often
appears a stronger and hence more pragmatic constraint.
One makes the additional Ansatz
that the condition (\ref{strict}) holds not only for the
particular value $y=y_1$, but for \emph{all} values of $y$, with
the value $N$ replaced by some function $f(y)$. In other words,
the Ansatz says that the Einstein equation implies no preferred
brane positions. This Ansatz implies that \bea \label{factor}
  n(t,y) &=& n_0(t)f(y) \nonumber \\
  a(t,y) &=& a_0(t)f(y) \nonumber \\
  b(t,y) &=& b(t,y) \ .
\eea

Substituting (\ref{factor}) into the $ty$-component of the
Einstein equation (\ref{einstein1}) and assuming $T_{ty}=0$, we
obtain the condition
\be
  f'\bdot =0 \ .
\ee

For a nontrivial warp factor, the above equation can only be
satisfied if $\bdot=0$. One can then redefine the coordinates $t$ and
$y$ to set the functions $n_0(t)$ and $b(y)$ to unity, so that the
metric reads \bea \label{factor2}
  n(t,y) &=& f(y) \nonumber \\
  a(t,y) &=& a_0(t)f(y) \nonumber \\
  b(t,y) &=& 1 \ .
\eea

As an aside, we note that the redefinition of the $y$-coordinate to obtain
$b=1$ contains a possible problem: if the original function $b(y)$ has zeros,
the required coordinate transformation may be singular. In particular, this
singularity may map either or both of the boundaries of the $y$-coordinate
from finite values to infinity, making the fifth dimension
non-compact. This issue was discussed in \cite{Lukas:1999}, and
singularities in the RS scenario have more generally been
considered in \cite{ Kaloper:1999, Gremm:2000, Arkani-Hamed:2000}.
In the present work, we simply assume that the metric is non-singular so that
the fifth dimension remains compact in such transformations.

We see that the factorized metric (\ref{factor2}) and the condition
$\bdot=0$ can be motivated by requiring Randall-Sundrum-type
solutions. If there is a weak time-dependence in $b$, we might
expect to approximately recover the factorizable metric
(\ref{factor2}), as noted in \cite{Lesgourges:2000}. It is known
that the metric (\ref{factor2}) does not allow for brane matter
(other than the cosmological constant) \cite{Mohapatra:2000, Lesgourges:2000}.
Next, we discuss the constraints for adding brane matter.

\subsection{Constraints on brane matter}

In addition to presenting a mechanism to generate
the hierarchy, the RS model had
another remarkable feature: the localization of
gravity into the vicinity of the Planck brane.
Clearly, one would like to preserve both of these features while
adding matter onto the TeV brane. Are there then any constraints on
possible equations of state for brane matter, and what
features of the model are responsible for them? We will next focus
on isolating the potential sources for trouble and find out what features
the model should have in order to accommodate generic brane matter.

The localization of gravity is easiest to
achieve in the orbifold geometry where the branes have a zero
thickness\footnote{Branes with finite thickness have been
considered in \cite{Kanti:1999a, Kanti:1999b, Gremm:2000}.
}. This assumption will be adopted in the present paper:

\begin{enumerate}

\item[4.] Both branes have zero thickness.

\end{enumerate}

\paragraph{Assumption 4.}
According to the above assumption, the stress tensor of brane
matter is proportional to a delta function. Thus the Einstein
tensor must contain delta functions as well.
The orbifolding of the $y$-coordinate introduces discontinuities
in the first $y$-derivatives of the metric, and delta functions in
the second $y$-derivatives. Since the metric is assumed to be
continuous, the only possible delta function contributions to the
Einstein tensor come from these second $y$-derivatives, $a''$ and
$n''$. This severely restricts the input of the metric to the
brane stress tensor (or vice versa).

The brane part of the Einstein equation (\ref{einstein1}) reads
\bea \label{einsteindelta}
  3\frac{1}{b^2}\frac{a''}{a}\bigg|_\delta &=& \kappa^2\left.
T^t_{\ t}\right|_{brane} \nonumber \\
  \frac{1}{b^2} \left(2\frac{a''}{a}+\frac{n''}{n}\right)\bigg|_\delta
&=& \kappa^2\left. T^i_{\ i}\right|_{brane} \ .
\eea

The notations $\delta$ and $brane$ refer to the delta function
parts of the derivatives and the stress tensor, respectively. With
the assumption of spatial homogeneity and isotropy the general
form of the brane stress tensor is \bea \label{stressdelta}
  \left. T^A_{\ \ B}\right|_{brane} =
\sum_{m=1,2} \frac{\delta(y-y_m)}{b(t,y_m)} {\rm diag }
(-\rho_m(t), p_m(t), p_m(t), p_m(t), 0) \ ,
\eea

\noindent where the index $m$ enumerates the branes, and we take
$y_2>y_1$. On the other hand, the delta function part of the
metric is related to the jumps of the first derivatives of the
metric, which can in turn be expressed in terms of the continuous
part of the metric \cite{Binetruy:1999a}:
\bea \label{deltacont}
  a''\big|_\delta &=& \sum_{m=1,2}\delta(y-y_m)[a'] \nonumber \\
&=& \sum_{m=1,2}\delta(y-y_m)(-1)^{m+1} 2\, a'_c \ ,
\eea

\noindent where $[a']$ is the discontinuity of $a'$,
$[a'(y)]:=\lim_{\varepsilon\rightarrow0}
(a'(y+\varepsilon)-a'(y-\varepsilon))$, and $a'_c$ is the
continuous part of $a'$. An identical relation holds for $n$.
Putting together (\ref{einsteindelta}), (\ref{stressdelta}) and
(\ref{deltacont}), we have \bea
  \label{branematter1} \pm \frac{2}{b}\frac{a'_c}{a}\bigg|_{y=y_m}
&=& -\frac{\kappa^2}{3}\rho_m \\
  \label{branematter2} \pm \frac{2}{b}\frac{n'_c}{n}\bigg|_{y=y_m}
&=& \frac{\kappa^2}{3} (3 p_m + 2\rho_m) \ .
\eea

\paragraph{The static configuration of branes.}
In the simplest scenario to generate the hierarchy between the
branes, one assumed that the warp factor $n$ is independent of the
cosmic time $t$, so that $n=f(y)$. In fact, there is always a
coordinate system where this is true. Any two-dimensional Riemannian manifold
is conformally flat, so we can change to conformal coordinates in the
($t,y$)--subspace. In conformal coordinates, the metric (\ref{metric1}) reads
\be \label{metric1b}
  ds^2 = -\tilde b(\tilde t,\tilde y)^2 d\tilde t^{\, 2}
+ \tilde b(\tilde t,\tilde y)^2 d\tilde y^2
+ \tilde a(\tilde t,\tilde y)^2\sum^3_{i=1}(dx^i)^2 \ .
\ee

We then orbifold in the $\ytilde$-direction, so that we have
two parallel branes at the fixed points at $\ytilde_1,\ytilde_2$.
Introducing the assumption $d\tilde b/d\tilde t=0$, the metric reduces to
\bea \label{metric2a}
  ds^2 = -\tilde b(\tilde y)^2 d\tilde t^{\, 2}
+ \tilde b(\tilde y)^2 d\tilde y^2
+ \tilde a(\tilde t,\tilde y)^2\sum^3_{i=1}(dx^i)^2 \ .
\eea

We redefine the $\tilde y$-coordinate so as to render the
metric into the form
\bea \label{metric2b}
  ds^2 = - f(\ytilde )^2 d\ttilde^2 + d\ytilde^2
+ a(\ttilde ,\ytilde )^2\sum^3_{i=1}(dx^i)^2 \ .
\eea

We call the above bulk-brane configuration the {\em static} configuration,
since the metric components in the $(\ttilde , \ytilde )$-subspace
are time-independent, and the branes are not moving. Note it was crucial to
introduce conformal coordinates \emph{before} orbifolding. Had we orbifolded
first and moved to conformal coordinates afterwards, the metric would
still have the form (\ref{metric2b}), but the branes would move through
the bulk instead of staying at fixed positions.



What are the constraints for brane matter in the static configuration?
If $a(\ttilde,\ytilde)=a_0(\ttilde )f(\ytilde )$, the only possible
equation of state is that of a cosmological
constant, $p_m =-\rho_m$. Giving up factorizability of $a(\ttilde , \ytilde )$
does not allow much more freedom. In that case
(\ref{branematter1}) allows $\rho_m$ to be time-dependent, but
because the {\em l.h.s} of (\ref{branematter2}) is time-independent,
the time-dependent part of $3p_m+2\rho_m$ must cancel. That
corresponds to a time-dependent density of two-dimensional domain
walls moving at the speed of light on the brane.
So the most
general equation of state is
\bea \label{constraint}
  \rho_m(\ttilde ) = \rho_{m(wall)}(\ttilde )+\rho_{m(vacuum)}
= -\frac{3}{2}p_{m(wall)}(\ttilde )-p_{m(vacuum)} \ .
\eea

In Section 3 we will discuss constraints for bulk matter.
Interestingly, we will find that the presence of an ideal fluid in
the bulk will force the bulk metric to be that of a static
configuration.


Several RS-type solutions
containing brane matter with arbitrary equations of state have been
presented \cite{Csaki:1999, Kanti:1999a, Kanti:1999b, Kanti:2000b,
Mohapatra:2000, Lesgourges:2000, Binetruy:1999a,Shiromizu:2000wj,
Binetruy:1999b}, to mention a few. First of all, it is an easier task to find
realistic brane cosmologies in a {\em single}-brane
configuration. Then one is not addressing the hierarchy problem, and
consequently there is more freedom in setting up the model.
For example, the brane could be moving across the bulk
geometry \cite{Kraus:1999it, Ida:1999ui, Mukohyama:2000wi},
or one can take the $n(t,y)$ factor in the metric to
be in a non-factorized form.
For two-brane systems, inclusion of generic matter requires
non-static configurations of branes.

We will next briefly discuss the non-static configurations with two branes.

\paragraph{Non-static configurations.} The strong constraints for
the brane matter followed because we introduced conformal coordinates
prior to orbifolding. Let us now see what happens if we orbifold in general
Gaussian normal coordinates. We start again from the general form of the metric
(\ref{metric1}), with $c=0$ and $K=0$. We then move to the orbifold geometry,
with parallel branes at the fixed points $y_1<y_2$. Further, we again
require that $\bdot=0$ and set $b=1$. So, we are dealing with the
following class of metrics
\be
  ds^2 = -n(t,y)^2 dt^2 + dy^2 + a(t,y)^2 \sum^3_{i=1} (dx^i)^2 \ ,
\ee

\noindent with a compact y-direction $y_1 \leq y \leq y_2$.
%

It is known that if $n(t,y)$ takes a non-factorized form, there
are single-brane solutions allowing generic matter.

The reference \cite{Kanti:2000b} presented a two-brane
configuration with generic matter on the TeV brane, while
satisfying the condition (\ref{strict}) up to leading order.
However, the fifth dimension was required to be static only
up to leading order, at higher orders the $b$-coefficient in
the metric may contain time-dependent terms.
%


\paragraph{Time-dependence of b.}

We discussed above that if one assumes $b$ to be fixed,
whether to avoid inducing
time-dependence in particle masses or for some other reason,
and have found that one easily ends up with overly constrained brane matter.

Perhaps the most viable option
for including brane matter is to allow for a time-dependent $b$.
It has been argued that time dependence in $b$ leads to
time-varying particle masses \cite{Csaki:1999}. Then any change in
$b$ must be negligible by the time of nucleosynthesis so as not to
contradict observation. As an aside, we note that these arguments
have depended on a specific form of the metric, and that it is not
obvious that a time-dependent $b$ would in general lead to such
time-dependence of $n$ and $a$ as to affect particle masses.

We should thus consider the possibility that
$b$ is almost stabilized but varies in time slowly enough not to
conflict with observation, either asymptotically approaching a
constant value or oscillating about a minimum. A $b$ varying
slowly should allow the inclusion of
realistic brane matter. The time-independent $b$ can then be
considered an approximation, and perhaps an asymptotic limit, of
this scenario. Treating time-dependent brane matter as a
perturbation against a background of bulk matter (including
possibly a cosmological constant) and brane cosmological constants
is supported by the fact that energy density of the universe from
the time of nucleosynthesis onwards is quite small in natural
units, even if the scale of the five-dimensional gravitation is in
the TeV range. Also, it seems in general more plausible that
dynamical evolution would lead to a $b$ with weak time-dependence
rather than a completely fixed $b$.

We now move on to discuss constraints for bulk matter.
Note that we will
continue to take $b$ to be constant, and assume that a
weak time-dependence (along with brane matter)
could be included as a perturbation. This line of thought has
been pursued in \cite{Csaki:1999, Grimstein:2000}.

\section{The Einstein Equation with Ideal Fluid}

\subsection{The stress tensor}

After general investigations of the metric and brane matter, we
now proceed to study the particular case of a bulk with ideal
fluid and a cosmological constant.

Bulk matter, either in the form of scalar fields or an ideal
fluid (which can sometimes be interpreted as a scalar field, and
vice versa) has been considered in many papers. However, the
scalar field studies, of which we mention only a few, have
concentrated on fixing the size of the fifth dimension
\cite{Goldberger:1999uk, Cline:2000, DeWolfe:1999}, inflation
\cite{Lukas:1999}, the cosmological constant problem
\cite{Arkani-Hamed:2000}, singularities and the
AdS/CFT-correspondence \cite{Gremm:2000}, or on more general
aspects of the formalism \cite{DeWolfe:1999}, not on obtaining
cosmological solutions. In papers of a more cosmological nature,
often only the $yy$-component of the bulk stress tensor has been
allowed to deviate from a cosmological constant \cite{Kanti:2000a,
Kanti:1999a, Kanti:1999b, Kanti:2000b, HBKim:2000}, and in any
case the $yy$-component has been taken to be different from the
$x^i x^i$-components \cite{Kim:2000, Mohapatra:2000,
Kennedy:2000}\footnote{\cite{Kennedy:2000} is particularly
interesting in that the $yy$- and $x^i x^i$-components
are initially assumed to be different but
are forced to be equal by the Ansatz used.},
so that the bulk matter cannot be interpreted as an ideal fluid.
Furthermore, complete and explicit cosmological solutions
for the case of a bulk stress tensor with non-trivial $tt$- and
$x^i x^i$-components are rarely presented; \cite{Grimstein:2000}
mentions one in passing. We will now present one such solution.

We take the branes and the bulk to contain some ideal fluids of
unspecified nature. For the moment, we do not make any assumptions
about the metric. It is clearest to introduce a local
orthonormal frame to find the form of the stress tensor. We
introduce coordinates $(\xhat^{\Ahat})=(\that,\xhat^{\ihat},\yhat)$
such that locally the five-dimensional line element takes the
form of the 5-d Minkowski metric,
\be
  ds^2 = -d\that^2 + \sum^3_{\ihat=1} (d\xhat^{\ihat})^2 + d\yhat^2 \ .
\ee

In the local orthonormal frame, the stress tensor for brane (bulk)
ideal fluids must be homogeneous and isotropic in the three (four)
spatial dimensions on the branes (in the bulk). We thus have \bea
  T^{\Ahat}_{\ \ \Bhat} &=& \sum_{m=1,2}
  \delta(\yhat-\yhat_m) {\rm diag }(-\rho_m, p_m, p_m, p_m,0) \nonumber \\ \mbox{}
  &&\ + {\rm diag}(-\rho-\Lambda, p-\Lambda, p-\Lambda, p-\Lambda, p-\Lambda )  \ .
\eea

\noindent In the above, $\rho_m, p_m$ are the energy densities and
pressures of ideal fluid on the two branes located at
\be
  \yhat_1 = \int^{y_1}_0 dy' b(t,y') \ ; \ \yhat_2 = \int^{y_2}_0 dy' b(t,y') \ ,
\ee

\noindent $\rho, p$ are the energy density and pressure of the
bulk ideal fluid and $\Lambda$ is the bulk cosmological constant
which we have for convenience separated out. In particular, note
that in the local orthonormal frame the pressure of an ideal fluid
in the $\yhat$ direction is {\em equal} to the pressure in the
$\xhat^{\ihat}$ directions. We now introduce the assumption of
homogeneity and isotropy in the directions parallel to the brane.
Then the pressures and energy densities cannot depend on the
coordinates $\xhat^{\ihat}$, only on the time $\that$ and, in the
case of the bulk fluid, the perpendicular direction $\yhat$. Thus,
expressing the stress tensor in general coordinates $x^A$, we have
\bea \label{stress}
  T^A_{\ \ B} &=& \sum_{m=1,2} \frac{\delta (y-y_m)}{b(t,y_m)}{\rm diag }
(-\rho_m(t), p_m(t), p_m(t), p_m(t), 0) \nonumber \\
  \mbox{} && + {\rm diag} (-\rho(t,y)-\Lambda, p(t,y)-\Lambda,
p(t,y)-\Lambda, p(t,y)-\Lambda, p(t,y)-\Lambda ) \ . \eea

\noindent The bulk ideal fluid is assumed to satisfy a linear
equation of state, \bea \label{state}
  p(t,y) &=& w \rho(t,y) \ .
\eea

We could for generality write the bulk ideal fluid as a sum of
components with different $w$, but since it turns out that $w$ can
take only one value, we prefer not to clutter the notation. Also,
one could in principle allow the coefficient $w$ to depend on $t$
and $y$, describing a time- and coordinate-dependent (that is,
interacting) mixture of ideal fluids. However, in this paper we
shall assume that $w$ is constant.

\subsection{Bulk ideal fluid implies a static configuration}

We again start from the metric (\ref{metric1}), with $c=0$ and $K=0$.
Before tackling the Einstein equation, we simplify the problem by taking
advantage of the conservation law of the stress tensor,
\be
  D_A T^A_{\ \ B} = 0 \ ,
\ee

\noindent which implies that matter on the branes and in the bulk satisfies the
following equations:
\bea
  \label{brane} \dot\rho_m
+ 3(\rho_m + p_m)\frac{\adot}{a}\bigg|_{y=y_m} &=& 0 \\
  \label{bulk1} \dot\rho + (\rho + p)
  \left( 3 \frac{\adot}{a} + \frac{\bdot}{b} \right) &=& 0 \\
  \label{bulk2} p' + (\rho + p) \frac{n'}{n} &=& 0 \ .
\eea

Interestingly, (\ref{bulk2}) implies that it is not
possible to have bulk ideal fluid with $p=0$ (assuming that the
warp factor is non-trivial). So, the conservation law of the
stress tensor already restricts the possible equations of state for the
bulk ideal fluid. We will shortly see that the full Einstein
equation (with the assumption $\bdot=0$) permits
only one particular equation of state (aside from the equation of
state of a cosmological constant), that of the stiff ideal fluid.

Substituting the equation of state (\ref{state}), we can integrate
the equations (\ref{bulk1}) and (\ref{bulk2}) for the bulk fluid to
obtain (for $w\neq0$)
\bea \label{rho}
  \rho(t,y) &=& \rho_0\, B(t)^{\frac{1+w}{w}}\, n(t,y)^{-\frac{1+w}{w}} \nonumber \\
  \rho(t,y) &=& \rho_0\, C(y)^{-\frac{1+w}{w}}\, a(t,y)^{-3(1+w)}\, b(t,y)^{-(1+w)} \ ,
\eea

\noindent where $\rho_0$ is a constant and $B$ and $C$ are some unknown
functions, with the powers of $w$ introduced for convenience. Excepting
the case of a cosmological constant, $w=-1$, we can rewrite
the equations (\ref{rho}) as (we introduce the assumption $\bdot=0$)
\bea \label{nanda1}
  \rho(t,y) &=& \rho_0\, B(t)^{\frac{1+w}{w}}\, n(t,y)^{-\frac{1+w}{w}} \nonumber \\
  n(t,y) &=& B(t)\, C(y)\, a(t,y)^{3w} \ .
\eea

The second equation in (\ref{rho}) is a relation between the metric functions
$n$ and $a$. On the other hand, the $ty$-component of the Einstein equation
also gives a relation between $n$ and $a$. According to (\ref{adoteqn}),
the assumption $T_{ty}=0$ implies (with $\bdot=0$)
\bea \label{nanda2}
  n(t,y) = A(t)^{-1} \adot(t,y) \ .
\eea

Combining the conservation law relations (\ref{nanda1}) and the no-flow relation (\ref{nanda2}), we have
\bea \label{nanda3}
  \rho(t,y) &=& \rho_0\, B(t)^{\frac{1+w}{w}}\, n(t,y)^{-\frac{1+w}{w}} \nonumber \\
  n(t,y) &=& B(t)\, C(y)\, a(t,y)^{3w} \nonumber \\
  \adot(t,y) &=& A(t)\, B(t)\, C(y)\, a(t,y)^{3w} \ .
\eea

We can integrate the equation for $a$ in (\ref{nanda3}) and obtain
$n$, $a$ and $\rho$ in terms of two unknown functions of $t$ and two
unknown functions of $y$. The form of these functions can be found in the
Appendix.

Let us now consider the Einstein equation (\ref{einstein1}) in the bulk,
with $\bdot=0$ (and $b=1$). Since the Einstein equation is local and the
metric is assumed to be continuous, the branes contribute only to boundary
conditions. Hence we can ignore the brane contribution to the stress tensor in
local bulk calculations. With the relation (\ref{nanda2}), the $ty$-component
of the Einstein equation is satisfied trivially. The remaining components of
(\ref{einstein1}) with the stress tensor (\ref{stress}) and the equation
of state (\ref{state}) read, away from the branes:
\bea \label{e2}
  \frac{a''}{a}+ \frac{a'^2}{a^2} - \frac{1}{n^2}\frac{\adot^2}{a^2}
&=& -\rhohat-\Lamhat \nonumber \\
  2\frac{a''}{a}+\frac{n''}{n}+\frac{a'}{a}
\left( \frac{a'}{a}+2\frac{n'}{n} \right)
-\frac{1}{n^2}\left[2\frac{\addot}{a}+\frac{\adot}{a}
\left(\frac{\adot}{a}-2\frac{\ndot}{n}\right)
\right] &=& 3(w\rhohat-\Lamhat) \nonumber \\
  \frac{a'}{a} \left(\frac{a'}{a} +\frac{n'}{n}\right)
-\frac{1}{n^2}\left[\frac{\addot}{a} + \frac{\adot}{a} \left(
\frac{\adot}{a} -\frac{\ndot}{n} \right)\right] &=& w\rhohat-\Lamhat \ ,
\eea

where
\be
  \rhohat := \frac{\kappa^2}{3} \rho \ , \ \Lamhat := \frac{\kappa^2 }{3} \Lambda \ .
\ee

Taking linear combinations of the above equations, we obtain the
following equivalent set:
\bea
  \label{e3a} -\frac{a''}{a}+\frac{n''}{n} -
2\frac{a'}{a}\left(\frac{a'}{a}-\frac{n'}{n}\right)
+\frac{2}{n^2}\left(-\frac{\addot}{a}+\frac{\adot}{a} \left(
\frac{\adot}{a} +\frac{\ndot}{n} \right)\right) &=& 3 (w+1) \rhohat \\
  \label{e3b} 3\frac{a''}{a}+\frac{n''}{n} &=& (w-1)\rhohat - 2\Lamhat \\
  \label{e3c} 2\frac{a''}{a}+\frac{n''}{n} - \frac{a'}{a}\left(2\frac{a'}{a}+\frac{n'}{n}\right) + \frac{1}{n^2}\left(\frac{\addot}{a}+\frac{\adot}{a}\left(2\frac{\adot}{a}-\frac{\ndot}{n} \right)\right) &=& 0 \ .
\eea

Of the above equations, (\ref{e3b}) is remarkable in that it does not contain
derivatives with respect to $t$. Thus, plugging in $n$, $a$ and $\rho$, which
we have expressed in terms of functions of $t$ and functions of $y$, into
(\ref{e3b}), we obtain an \emph{algebraic} equation involving functions of
$t$. The details are again relegated to the Appendix. From this algebraic
equation we see that $n$ and $a$ have the following factorizable
form\footnote{The special case $w=-\frac{1}{3}$ would yield a
non-factorizable metric. However, this requires $\Lambda=0$, so the metric
does not have exponential warping and is of no interest as regards the
hierarchy problem.}:
\bea
  n(t,y) &=& n_0(t) f(y) \nonumber \\
  a(t,y) &=& a_0(t) f(y) \ .
\eea

Shifting to cosmic time, we arrive at the simple factorizable metric
(\ref{factor2}):
\bea \label{factor2b}
  n(t,y) &=& f(y) \nonumber \\
  a(t,y) &=& a_0(t) f(y) \ .
\eea

So, the presence of bulk ideal fluid leads (given our four assumptions)
to the static configuration with a simple factorizable metric.

\subsection{Solution of the Einstein equation}

We have shown that when the bulk contains ideal fluid, our four assumptions
lead to the factorizable metric (\ref{factor2b}). We now complete the solution
of the Einstein equation with bulk and brane ideal fluids (of course, the
brane ideal fluids reduce to cosmological constants, as we see from
(\ref{branematter1})).

With the factorizable metric (\ref{factor2b}), the equations
(\ref{e3a}) to (\ref{e3c}) read
\bea
  \label{e4a} \rhohat &=& \frac{2}{3(w+1)}
  \left(\left(\frac{\adot_0}{a_0}\right)^2 - \frac{\addot_0}{a_0} \right) f^{-2} \\
  \label{e4b} \frac{f''}{f} &=& \frac{w-1}{4}\rhohat - \frac{\Lamhat}{2} \\
  \label{e4c} f''f-(f')^2 &=& -\frac{1}{3}\frac{\addot_0}{a_0}
- \frac{2}{3}\left(\frac{\adot_0}{a_0}\right)^2 \equiv -\frac{\rm C}{9} \ .
\eea

In the last equation we have introduced a constant C. The {\em
l.h.s.} of (\ref{e4b}) is independent of $t$, while from
(\ref{e4a}) we see that $\rhohat$ depends\footnote{The possibility
$\frac{\adot_0^2}{a_0^2} - \frac{\addot_0}{a_0} = \textrm{constant}$
would reduce the bulk fluid to a cosmological constant.} on $t$.
Thus, we obtain the result $w=1$\footnote{This result
appeared in \cite{Kennedy:2000} in a different setting,
valid only in the case $\bdot\neq0$.}. In other words, the only allowed
equation of state for the bulk ideal fluid is
\be \label{w1}
  p = \rho \ .
\ee

This result does not depend on the choice $K=0$. We postpone
discussion of the properties of an ideal fluid with this equation
of state to section 3, and proceed to find the corresponding
cosmological solutions.

With $w=1$, the equations (\ref{e4a}) to (\ref{e4c}) read
\bea
  \label{e5a} \rhohat &=& \frac{1}{3}\left(\left( \frac{\adot_0}{a_0}\right)^2
  -\frac{\addot_0}{a_0} \right) f^{-2} \\
  \label{e5b} \frac{f''}{f} &=& -\frac{\Lamhat}{2} \\
  \label{e5c} f''f - (f')^2 &=& -\frac{\rm C}{9} \\
  \label{e5d} \frac{\addot_0}{a_0} + 2 \left(\frac{\adot_0}{a_0}\right)^2
&=& \frac{\rm C}{3} \ .
\eea

For $K\neq0$, the time dependence of $\rhohat$ and the equation
for $a_0$ would change, but the equations for the warp factor
would remain unchanged. Note that the bulk cosmological
constant contributes only to the warp factor, not to the
cosmological expansion factor $a_0$. Interestingly,
the bulk fluid accumulates at the brane where $f$ has its minimum
value, in other words at the TeV brane, a result noted in \cite{Kim:2000}.

\paragraph{Comparison with the FRW model.} It is interesting to compare the
equations (\ref{e5a}) and (\ref{e5d}) with the corresponding
equations for the $K=0$ Friedmann-Robertson-Walker model. We take
the same matter content for the FRW model: an ideal fluid with the
equation of state $p_{(3)}=\rho_{(3)}$ and a cosmological constant
$\Lambda_{(3)}$ (we use the subscript 3 as a reminder that there
are only three spatial dimensions). The Einstein equation for the
FRW model with this matter content can be written as
\bea
  \label{FRWa} \rhohat_ {(3)}&=& \frac{1}{3}
  \left(\left(\frac{\adot}{a}\right)^2-\frac{\addot}{a}\right) \\
  \label{FRWb} \frac{\addot}{a} + 2 \left(\frac{\adot}{a}\right)^2
&=& 3\Lamhat_{(3)} \ .
\eea

Comparison of (\ref{e5a}) and (\ref{e5d}) with (\ref{FRWa}) and
(\ref{FRWb}) shows that $a_0$ in the RS case is exactly the same
as $a$ in the FRW case, with C playing the role of a cosmological
constant, $\rm C = 9\Lamhat_{(3)}$, and that $\rhohat$ equals
$\rhohat_{(3)}$, save for the factor $f^{-2}$. These results hold
also for $K\neq0$. So, bulk ideal fluid in the RS model behaves
exactly like ideal fluid in the FRW model, modulo the warp factor.
However, we will shortly see that in the RS case only the sign of
C has any physical meaning, in contrast to the FRW case, where the
magnitude of $\Lamhat_{(3)}$ sets the timescale of the universe.

We analyze the cases $\rm C=0$ and $\rm C\neq0$ separately.

\paragraph{$\bfm{C=0}$.}

The equations (\ref{e5a}) to (\ref{e5d}) now read \bea
  \rhohat &=& \frac{1}{3}\left(\left(\frac{\adot_0}{a_0}\right)^2
-\frac{\addot_0}{a_0}\right) f^{-2} \\
  \frac{f''}{f} &=& -\frac{\Lamhat}{2} \\
  f''f - (f')^2 &=& 0 \\
  \frac{\addot_0}{a_0}+2\left(\frac{\adot_0}{a_0}\right)^2 &=& 0 \ .
\eea

The above equations have the following unique\footnote{Apart from
the trivial solution $\adot_0=0$, which would lead to $\rho=0$ and
the original RS model.} solution \bea
  \label{a1} a_0(t) &=& \left(\frac{t}{\tau}\right)^{1/3} \\
  \label{f1} f(y) &=& f_0 e^{\pm y\sqrt{-\Lamhat/2}} \\
  \label{rho1} \rho(t,y) &=& \frac{1}{3\kappa^2}\frac{1}{t^2} f(y)^{-2} \ .
\eea

We have normalized by $a_0(0)=0$. The bulk cosmological constant
has to be negative, $\Lamhat<0$.

The solution contains two free parameters, $\tau$ and $f_0$. A
choice of the time parameter $\tau$ corresponds to choosing the
unit of time, so that $\tau$ has no physical meaning, as in the
FRW case. A choice of the warp parameter $f_0$ corresponds to
choosing the origin of the $y$-coordinate. The metric and the
energy density are invariant under the scaling
$t\rightarrow\lambda t$, $x^i\rightarrow\lambda x^i$,
$f_0\rightarrow\lambda^{-1} f_0$, so that $f_0$, or the placement
of the origin, has no physical meaning.

The $\rm C=0$ solution contains no physical degrees of freedom
other than the value of the bulk cosmological constant: the metric
and the bulk energy density (and, as we will see, the brane
cosmological constants) are fixed once the bulk cosmological
constant is specified. This suggests that the solution is
unstable, possibly collapsing to one of the $\rm C\neq0$ solutions
when perturbed.

\paragraph{$\bfm{C\neq 0}$.}

In this case the equations (\ref{e5a}) to (\ref{e5d}) read
\bea
  \label{6a} \rhohat &=& \frac{1}{3}\left(\left(\frac{\adot_0}{a_0}\right)^2
-\frac{\addot_0}{a_0}\right) f^{-2} \\
  \frac{f''}{f} &=& -\frac{\Lamhat}{2} \\
  f''f - (f')^2 &=& -\frac{\rm C}{9} \\
  \label{6b} \frac{\addot_0}{a_0}+2\left(\frac{\adot_0}{a_0}\right)^2 &=& \frac{\rm C}{3} \ .
\eea

The above equations have the solution\footnote{The equation for
$a_0$ allows de Sitter and anti-de Sitter solutions when $\rm
C>0$, but these would again give $\rho=0$. These solutions were
first presented in \cite{Kaloper:1999, Nihei:1999}.} \bea
  \label{a2} a_0(t) &=& a_1\left\{ \begin{array}{l}
\sin^{1/3}(\sqrt{\rm{|C|}}~t) \ \qquad \rm C<0  \\
\sinh^{1/3}(\sqrt{\rm C}~t) \ \qquad \rm C>0 \end{array} \right.\\
  \label{f2} f(y) &=& f_1 e^{y\sqrt{-\Lamhat /2}}
  + \frac{\rm C}{18\Lamhat f_1} e^{-y\sqrt{-\Lamhat /2}} \\
  \label{rho2} \rho(t,y) &=& \frac{\rm{|C|}}{3\kappa^2}
   f(y)^{-2}\left\{ \begin{array}{l}
\sin^{-2}(\sqrt{\rm{|C|}}~t)\ \qquad \rm C<0  \\
\sinh^{-2}(\sqrt{\rm C}~t)\ \qquad \rm C>0 \end{array} \right. \ .
\eea

We have again used the normalization $a_0(0)=0$. The equations for
$f$ would allow a positive (zero) bulk cosmological constant,
resulting in trigonometric functions (a linear function), but in
order to solve the hierarchy problem we take $\Lamhat<0$.

As in the previous case, the parameter $a_1$ has no physical
meaning, and can be set to unity, while $f_1$ corresponds to the
choice of the origin of $y$. Rewriting $f_1$ in terms of a
coordinate value $y_0$, we have \bea
  \label{a2b} a_0(t) &=& \left\{ \begin{array}{l}
\sin^{1/3}(\sqrt{\rm{|C|}}~t) \ \qquad \rm C<0 \\
\sinh^{1/3}(\sqrt{\rm C}~t)  \ \qquad \rm C>0 \end{array} \right.
\\
  \label{f2b} f(y) &=& \sqrt{\frac{2{\rm{|C|}}}{9|\Lamhat|}}\left\{ \begin{array}{l}
\cosh\big(\sqrt{|\Lamhat|/2}~(y-y_0)\big) \ \qquad \rm C<0 \\
\sinh\big(\sqrt{|\Lamhat|/2}~(y-y_0)\big) \ \qquad \rm C>0
\end{array} \right. \\
  \label{rho2b} \rho(t,y) &=& \frac{|\Lambda|}{2}\left\{ \begin{array}{l}
\cosh^{-2}\big(\sqrt{|\Lamhat| /2}~(y-y_0)\big) \
\sin^{-2}(\sqrt{\rm{|C|}}~t)\ \qquad \rm C<0  \\
\sinh^{-2}\big(\sqrt{|\Lamhat| /2}~(y-y_0)\big) \
\sinh^{-2}(\sqrt{\rm C}~t)\ \qquad \rm C>0 \end{array} \right. \ .
\eea

We see that the metric and the energy density are invariant under
the scaling $t\rightarrow\lambda t$, $x^i\rightarrow\lambda x^i$,
$\rm C\rightarrow\lambda^{-2}\rm C$. In other words, only the sign
of C has any physical meaning, the magnitude is irrelevant. In
particular, C does not introduce a new mass scale into the model.

Unlike in the case $\rm C=0$ (and the original RS proposal), the
model is \emph{not} invariant under translations of the
$y$-coordinate, so that the choice of origin of $y$ does have a physical
meaning. The translational invariance is broken by the
$y$-dependence in the bulk matter: $y_0$ is not a free parameter
but is set by the bulk energy density, as we see from
(\ref{rho2b}). There are two physical degrees of freedom: the
value of the bulk cosmological constant and the bulk energy
density. (In addition, there are of course the brane cosmological
constants, which, as we will see shortly, are also free
parameters.)

In the case $\rm C>0$ the function $f$ has a zero at $y=y_0$. The
bulk energy density and the scalar curvature diverge when $f=0$,
so the singularity at $y=y_0$ is physical, not an artifact of the
coordinate system we have chosen.  In what follows, we simply
avoid the singularity by constraining the value $y_0$ not to lie
between the brane positions in the case $\rm C>0$.

\subsection{Branes and fine-tuning}

Having solved the Einstein equation in the bulk, we now turn to
the branes, which, as noted earlier, provide boundary conditions.

\paragraph{Brane cosmological constants.} The factorizable metric implies
that the branes contain only cosmological constants. We switch to
the notation $\rho_m=-p_m=:\Lambda_m$. Inserting the brane
contribution to the stress tensor from (\ref{stress}) into the
Einstein equation (\ref{einstein1}), we get \bea \label{match}
  \frac{[f']}{f}\bigg|_{y=y_m} &=& -\Lamhat_m \ ,
\eea

\noindent where $y_1, y_2$ ($y_2 > y_1$) are the orbifold fixed
points, where the branes are located\footnote{In the case $\rm
C>0$ we demand $y_0 < y_1$ or $y_0 > y_2$ to avoid having a
singularity.}. As before, the notation $[f']$ refers to the
discontinuity of $f'$. Writing the discontinuity in terms of the
continuous part $f_c'$, we have \bea \label{match1}
  2\frac{f_c'}{f}\bigg|_{y=y_1} &=& -\Lamhat_1 \nonumber \\
  -2\frac{f_c'}{f}\bigg|_{y=y_2} &=& -\Lamhat_2 \ .
\eea

For $\rm C=0$ (and the original RS proposal) we have $f(y)=f_0
e^{-|y|\sqrt{|\Lamhat|/2}}$. Then (\ref{match1}) gives the result
$\Lamhat_1=-\Lamhat_2=\sqrt{2|\Lamhat|}$, an unfortunate
fine-tuning. However, for $\rm C<0$ we have instead \bea
\label{match2}
  \Lamhat_1 &=& -\sqrt{2|\Lamhat|}\tanh(\sqrt{|\Lamhat|/2}~(y_1-y_0)) \nonumber \\
  \Lamhat_2 &=& \sqrt{2|\Lamhat|}\tanh(\sqrt{|\Lamhat|/2}~(y_2-y_0)) \ ;
\eea

\noindent for $\rm C>0$ the hyperbolic tangent is replaced by a
hyperbolic cotangent\footnote{Note that either set of equations
fixes $y_2-y_1$, the size of the fifth dimension.}. The above
equations for the case $\rm C<0$ were found in \cite{Kanti:2000a}
in a slightly different setting. The values of hyperbolic tangent
lie in the interval from --1 to +1 (excluding $\pm1$), and the
values of hyperbolic cotangent stretch from minus infinity to plus
infinity, excluding the interval from --1 to +1. Thus, the ratio
of the brane cosmological constants to the square-root of the bulk
cosmological constant determines the value of C:
\bea
  \rm C &<& 0 \qquad \textrm{for }\ \frac{|\Lamhat_m|}
{\sqrt{2|\Lamhat|}}<1 \nonumber \\
  \rm C &=& 0 \qquad \textrm{for }\ \frac{|\Lamhat_m|}
{\sqrt{2|\Lamhat|}}=1 \nonumber \\
  \rm C &>& 0 \qquad \textrm{for }\ \frac{|\Lamhat_m|}
{\sqrt{2|\Lamhat|}}>1 \ .
\eea

In the present model, the Einstein equation implies no fine-tuning
of parameters, unlike in the original RS proposal. For a given
value of the bulk cosmological constant $\Lamhat$, the absolute
values of both brane cosmological constants have to be smaller
than, bigger than or equal to $\sqrt{2|\Lamhat|}$, but are
otherwise unrestricted by the Einstein equation. With two
equations to satisfy and two constants, $y_1-y_0$ and $y_2-y_0$,
at our disposal, no fine-tuning is needed to obtain a solution to
(\ref{match1}).

The signs of the brane cosmological constants are opposite for
$\rm C=0$ and $\rm C>0$. This can also be the case for $\rm C<0$.
However, for $\rm C<0$ it is also possible for both brane
cosmological constants to be positive: this requires the value of
$y_0$ to lie between the brane positions, $y_1 < y_0 < y_2$.
Furthermore, we have the interesting possibility of a \emph{zero}
cosmological constant on one brane and a positive cosmological
constant on the other: this requires one of the branes to be
placed at the point where $f$ has its minimum, $y_0=y_1$ or
$y_0=y_2$. (It is not possible to have two branes with
non-positive cosmological constants.) Thus, there is no need for
negative energy densities on the branes, essentially because the
warp factor is not a monotonical function, but has a minimum. This
was first noticed in \cite{Kaloper:1999} and further emphasized in
\cite{Kanti:2000a}. Of course, the bulk cosmological constant
still has to be negative to obtain an exponential warp factor.

\paragraph{The hierarchy problem.} The solution to the hierarchy problem
introduces a new equation, namely \bea \label{hier}
  \frac{f(y_1)}{f(y_2)}\sim 10^{16} \ .
\eea

For $\rm C<0$, the above equation reads \bea \label{large1}
  \frac{ \cosh(\sqrt{|\Lamhat|/2}~(y_1-y_0)) }
{ \cosh(\sqrt{|\Lamhat|/2}~(y_2-y_0)) } \sim 10^{16} \ ;
\eea

\noindent for $\rm C>0$ the hyperbolic cosine is replaced by a
hyperbolic sine. Combining (\ref{large1}) with (\ref{match2}), we
can write the hierarchy condition (\ref{hier}) as the following
equation, valid for all values of C: \bea \label{large2}
  \frac{\Lamhat_1^2}{2 |\Lamhat|}-1 \sim 10^{32}
\left(\frac{\Lamhat_2^2}{2 |\Lamhat|}-1\right) \ .
\eea

We see that the only way to avoid introducing unnaturally large
numbers is to fine-tune the cosmological constants, $\Lamhat_1^2 =
\Lamhat_2^2 = 2 |\Lamhat|$, corresponding to $\rm C=0$.

The above result is not dependent on the presence of bulk ideal
fluid: if we put $\rho=0$, our solutions disappear, but we get new
solutions with the same problems. For $\rm C=0$ we get a constant
$a_0$ (the original RS solution) and for $\rm C>0$ we get de
Sitter and anti-de Sitter solutions, found in \cite{Kaloper:1999, Nihei:1999};
the case $\rm C<0$ becomes disallowed. The warp
factor is unchanged by the absence of bulk ideal fluid (except
that $y_0$ becomes a free parameter), so the condition
(\ref{large2}) is also unchanged. This means that the fine-tuning
problem is $\emph{not}$ inherent to the Einstein equation.
Specifically, it is not due to the condition $\bdot=0$, as
sometimes claimed \cite{Csaki:1999}. It is rather a rephrasing of
the hierarchy problem. In this sense the RS model does not provide
quite a satisfactory solution to the hierarchy problem.

\section{Discussion}

As is well known, the RS solution is quite precarious; bulk and
brane sources for gravity must be chosen very carefully for the RS
solution to emerge from the Einstein equation. However, it turns
out that in two-brane systems one easily end up with severe
constraints for the sources; this fact is not apparent from
the literature. We hope that our results help to identify the
potential problem sources in trying
introduce realistic matter into the bulk or branes, while preserving
the main features of the RS scenario: the hierarchy generation
and the localization of gravity. The need to include generic
brane matter in particular creates problems in
trying to interpret our universe as a brane world.

In Section 2 we found that under minimal assumptions on the RS
model, a stabilized radion and zero thickness branes, one cannot
introduce realistic brane matter\footnote{In this context,
we mean homogeneous and isotropic matter with equations of state
corresponding to a combination of dust and radiation.} in static
two-brane configurations.
The only allowed equation of state, other than that of a
cosmological constant, corresponds to a fluid of two dimensional
domain walls moving at the speed of light. Even that requires
giving up factorizability of the bulk metric. Further relaxation
of the constraints for brane matter requires a time-dependent
and non-factorized warp factor, and most likely a slow
stabilization of the radion as the
branes settle to the final fixed proper distance. This idea is supported
by the result of Section 3, which shows that when the bulk is filled with
ideal fluid and the radion is completely stabilized, the warp factor is
necessarily factorizable. The constraint for the equation of state of the
bulk ideal fluid also points in this direction.

The dynamical stabilization of the radion degree of freedom can
be expected to be highly nontrivial. Nevertheless, unless for
some reason the universe started out with all the degrees of
freedom in the ground state, some dynamics should be expected. In
the context of cosmology, a natural initial condition for the
radion could be any value compatible with the uncertainty
principle, as in chaotic inflation. The radion potential energy
must then be released in some way. It is conceivable that the
extra radion energy is dissipated into purely gravitational
degrees of freedom, but there is no guarantee that a RS-type
solution would result; this remains to be studied. Dissipation into
fields living on the brane has been studied in \cite{Csaki:1999}.
The third possibility, radion decaying into bulk degrees of freedom, assumed
in this paper, is possible but highly constrained, as we have
shown. The only admissible equation of state for ideal fluid in
the bulk was found to be  $p = \rho \ $, representing the so-called
stiff ideal fluid. Here "stiff" reflects the fact that the
velocity of sound in the fluid is equal to the velocity of light.
Concretely, such a fluid corresponds to a classical free massless,
coherent scalar field (not to be confused with massless
radiation), something that was considered in a static setting in
\cite{Kanti:2000a}. Further, the massless bulk scalar field should
not be confused with the massive Goldberger-Wise bulk scalar
field. Whether such a field could be coupled to the radion is
beyond the scope of the present study.

The no-go flavor of our results is meant to clarify what are
the most promising directions for modifying the two-brane models
so that other equations of state for the brane or bulk
ideal fluids can be allowed.
In the case of the bulk, a natural guess would be to start with a
different Ansatz for the bulk geometry. This corresponds to a
different assumption about the spacetime symmetries. For example, if
the bulk geometry would correspond to an AdS black hole, one might
expect the equation of state for massless radiation to become allowed.

\section*{Acknowledgements}

KE was partly supported by the Academy of Finland under the
contract 101-35224. We would like to thank W. Goldberger, C. Montonen,
L. Randall, and M. Wise for some helpful discussions as we
were getting into this subject. We also thank J. Cline, A. Karch,
D. Langlois, T. Shiromizu, and their collaborators
for correspondence on the topic of constraints for brane matter.

\appendix
\section*{Appendix: Bulk Ideal Fluid and Factorization}

We present here details of the calculation of section 3.2 which shows that
the presence of bulk ideal fluid leads to a factorizable metric.  The
starting point was the equations (\ref{nanda3}):
\bea \label{A1}
  \rho(t,y) &=& \rho_0\, B(t)^{\frac{1+w}{w}}\, n(t,y)^{-\frac{1+w}{w}} \nonumber \\
  n(t,y) &=& B(t)\, C(y)\, a(t,y)^{3w} \nonumber \\
  \adot(t,y) &=& A(t)\, B(t)\, C(y)\, a(t,y)^{3w} \ .
\eea

Integrating the equation for $a$, we obtain the following expressions for
$n$, $a$ and $\rho$:
\bea \label{A2}
w=\frac{1}{3}&& \left\{ \begin{array}{lll}
  \rho(t,y) &=& \rho_0\, C(y)^{-4}\, e^{-4 (C(y)E(t)+D(y))} \\
  n(t,y) &=& B(t)\, C(y)\, e^{C(y)E(t)+D(y)} \\
  a(t,y) &=& e^{C(y)E(t)+D(y)}\ , \end{array} \right. \nonumber \\
w\neq\frac{1}{3}&& \left\{ \begin{array}{lll}
  \rho(t,y) &=& \rho_0\, C(y)^{-\frac{1+w}{w}} (C(y)E(t)+D(y))^{-3\frac{1+w}{1-3w}} \\
  n(t,y) &=& B(t)\, C(y)\, (C(y)E(t)+D(y))^{\frac{3w}{1-3w}} \\
  a(t,y) &=& (C(y)E(t)+D(y))^{\frac{1}{1-3w}}\ . \end{array} \right.
\eea

In the above expression, $D(y)$ is a new unknown function, and $E(t)$
is proportional to the integral of $A(t) B(t)$. We have now separated
the $t$- and $y$-dependence of the metric. The linear combination of
components of the Einstein equation which does not involve derivatives with
respect to $t$ is (\ref{e3b}):
\bea \label{A3}
  3\frac{a''}{a}+\frac{n''}{n} &=& (w-1)\rhohat - 2\Lamhat \ .
\eea

Now we simply insert (\ref{A2}) into (\ref{A3}). Let us consider the cases
$w=\frac{1}{3}$ and $w\neq\frac{1}{3}$ separately.

\paragraph{$\bfm{w=\frac{1}{3}}$.}

In this case (\ref{A3}) reduces to
\bea \label{A4}
  4 C'^2 E(t)^2 + \left(4 D'' + 8 C' D' + 2\frac{C'^2}{C}\right) E(t)
+\left(4 D''+ 4 D'^2 +\frac{A''}{A} + 2 \frac{C'}{C} D'\right) \nonumber \\
  = (w-1)\, \rhohat_0\, C^{-4} e^{-4 (C E(t)+D)} - 2\Lamhat \ .
\eea

It is impossible for the above equation to be satisfied at all times $t$ unless
$\dot E=0$ or $C=0$, both of which correspond to $\adot=0$, and thus static
cosmology.

\paragraph{$\bfm{w\neq\frac{1}{3}}$.}

Now (\ref{A3}) can be written as
\bea \label{A5}
  \frac{6 w (1+3 w)}{(1-3w)^2} D^2 \left(\frac{D'}{D}-\frac{C'}{C}\right)^2 (C E(t)+D)^{-2} \nonumber \\
  + \frac{3(1+w)}{1-3w}D\left(\frac{D''}{D}-\frac{C''}{C} + \frac{6 w}{1-3w}\frac{C'}{C}\left(\frac{D'}{D}-\frac{C'}{C}\right) \right) (C E(t)+D)^{-1} \nonumber \\
  + \frac{4}{1-3w}\left(\frac{C''}{C}+\frac{3 w}{1-3w}\frac{C'^2}{C^2} \right) \nonumber \\
= (w-1)\, \rhohat_0\, C^{-\frac{1+w}{w}} (C E(t)+D)^{-3\frac{1+w}{1-3 w}}
- 2\Lamhat \ .
\eea

The above equation is less transparent than (\ref{A4}). Unless
$w=-\frac{1}{3}$, there are three powers of $C E+D$ on the \emph{l.h.s.},
so by equating the coefficients of the powers of $C E+D$ we obtain three
equations. These equations can only be satisfied if $D=0$ (or $D\propto C$,
which amounts to the same thing by a redefinition of $C$). From (\ref{A2})
we see that this corresponds to a factorizable metric. This solution also
requires $w=1$, though for reasons of clarity we preferred to present
this result at a later stage in section 3.

As for the special case $w=-\frac{1}{3}$, the above equation then does have
solutions with a non-factorizable metric. However, a substitution to the full
Einstein equation (\ref{e3a})-(\ref{e3c}) shows that this requires $\Lambda=0$,
and as a result the metric does not have an exponential warp factor and is of
no interest as far as the hierarchy problem is concerned.


\begin{thebibliography}{99}

\bibitem{Randall:1999ee} {\normalsize Lisa Randall and Raman Sundrum.
\newblock {\it A large mass hierarchy from a small extra dimension}.
\newblock {\em Phys. Rev. Lett.}, 83:3370, 1999.
\newblock {\bf hep-ph/9905221}.
}

\bibitem{Randall:1999vf} {\normalsize Lisa Randall and Raman Sundrum.
\newblock {\it An alternative to compactification}.
\newblock {\em Phys. Rev. Lett.}, 83:4690, 1999.
\newblock {\bf hep-th/9906064}.
}

\bibitem{Csaki:1999} Csaba Cs\'{a}ki, Michael Graesser, Lisa
Randall and John Terning.
\newblock {\it Cosmology of brane models with radion stabilization}
\newblock {\bf hep-ph/9911406}.

\bibitem{Goldberger:1999uk} Walter D. Goldberger and Mark B. Wise.
\newblock {\it Modulus stabilization with bulk fields}.
\newblock {\em Phys. Rev. Lett.}, 83:4922, 1999.
\newblock {\bf hep-ph/9907447}.
\newblock {\it Bulk Fields in the Randall-Sundrum Compactification Scenario}.
\newblock {\bf hep-ph/9907218}.

\bibitem{Cline:2000} James M. Cline and Hassan Firouzjahi.
\newblock {\it Brane-world cosmology of modulus
stabilization with a bulk scalar field}.
\newblock {\bf hep-ph/0005235}.

\bibitem{DeWolfe:1999} O. DeWolfe, D.Z. Freedman, S.S. Gubser and A. Karch.
\newblock {\it Modeling the fifth dimension with scalars and gravity}.
\newblock {\bf hep-th/9909134}.

\bibitem{Kanti:2000a} Panagiota Kanti, Keith A. Olive and Maxim
Pospelov.
\newblock {\it Static solutions for brane models with a bulk scalar field}.
\newblock {\bf hep-ph/0002229}.

\bibitem{Luty:1999} Markus A. Luty and Raman Sundrum.
\newblock {\it Radius stabilization and anomaly-mediated supersymmetry breaking}.
\newblock {\bf hep-th/9910202}.

\bibitem{Steinhardt:1999} Paul J. Steinhardt.
\newblock {\it General considerations of the cosmological constant and
the stabilization of moduli in the brane-world picture}.
\newblock {\bf hep-ph/9907080}.

\bibitem{Kim:2000} Jihn E. Kim and Bumsok Kyae.
\newblock {\it Exact cosmological solution and modulus
stabilization in the Randall-Sundrum model with bulk matter}.
\newblock {\bf hep-th/0005139}.

\bibitem{Kanti:1999a} Panagiota Kanti, Ian. I. Kogan,
Keith A. Olive and Maxim Pospelov.
\newblock {\it Cosmological 3-brane solutions}.
\newblock {\bf hep-ph/9909481}.

\bibitem{Kanti:1999b} Panagiota Kanti, Ian I. Kogan, Keith A.
Olive and Maxim Pospelov.
\newblock {\it Single-brane cosmological solutions with a stable compact
extra dimension}
\newblock {\em Phys. Rev.}, D61:106004, 2000.
\newblock {\bf hep-ph/9912266}.

\bibitem{Kanti:2000b} Panagiota Kanti, Keith A. Olive and Maxim Pospelov.
\newblock {\it Solving the hierarchy problem in two-brane cosmological models}.
\newblock {\bf hep-ph/0005146}.

\bibitem{HBKim:2000} Hang Bae Kim.
\newblock {\it Cosmology of Randall-Sundrum models with an
extra dimension stabilized by balancing bulk matter}.
\newblock {\bf hep-th/0001209}.

\bibitem{Mohapatra:2000} R.N. Mohapatra, A. P\'erez-Lorenzana and C.S. de S. Pires.
\newblock {\it Cosmology of brane-bulk models in five dimensions}.
\newblock {\bf hep-ph/0003328}.

\bibitem{Kennedy:2000} Conall Kennedy and Emil M. Prodanov.
\newblock {\it Standard cosmology from sigma-model}.
\newblock {\bf hep-th/0003299}.

\bibitem{Kaloper:1999} Nemanja Kaloper.
\newblock {\it Bent domain walls as braneworlds}.
\newblock {\bf hep-th/9905210}.

\bibitem{Lukas:1999} Andr\'e Lukas, Burt A. Ovrut and Daniel Waldram.
\newblock {\it Boundary inflation}.
\newblock {\bf hep-th/9902071}.

\bibitem{Gremm:2000} Martin Gremm.
\newblock {\it Thick domain walls and singular spaces}.
\newblock {\bf hep-ph/0002040}.

\bibitem{Arkani-Hamed:2000} Nima Arkani-Hamed, Savas Dimopoulos,
Nemanja Kaloper and Raman Sundrum.
\newblock {\it A small cosmological constant from a large extra dimension}.
\newblock {\bf hep-th/0001197}.

\bibitem{Lesgourges:2000} Julien Lesgourges, Sergio Pastor,
Marco Peloso and Lorenzo Sorbo.
\newblock {\it Cosmology of the Randall-Sundrum model after dilaton stabilization}.
\newblock {\bf hep-ph/0004086}.

\bibitem{Binetruy:1999a} Pierre Bin\'etruy, C\'edric Deffayet and David Langlois.
\newblock {\it Non-conventional cosmology from a brane-universe}.
\newblock {\bf hep-ph/9905012}.

\bibitem{Shiromizu:2000wj} T. Shiromizu, K. Maeda and M. Sasaki.
\newblock {\it The Einstein equations on the 3-brane world}.
\newblock {\em Phys. Rev.}, D62:024012, 2000.
\newblock {\bf gr-qc/9910076}.

\bibitem{Binetruy:1999b} Pierre Bin\'etruy, C\'edric Deffayet,
Ulrich Ellwanger and David Langlois.
\newblock {\it Brane cosmological evolution in a bulk with cosmological constant}.
\newblock {\bf hep-th/9910219}.

\bibitem{Kraus:1999it} P. Kraus.
\newblock {\it Dynamics of anti-de Sitter domain walls}-
\newblock {\em JHEP} 9912, 011, 1999.
\newblock {\bf hep-th/9910149}.

\bibitem{Ida:1999ui} D. Ida.
\newblock {\it Brane-world cosmology}.
\newblock {\bf gr-qc/9912002}.

\bibitem{Mukohyama:2000wi} S. Mukohyama, T. Shiromizu and K.
Maeda.
\newblock {\it Global structure of exact cosmological solutions in the brane
world}.
\newblock {\em Phys. Rev.}, D62:024028, 2000.
\newblock {\bf hep-th/9912287}.


\bibitem{Grimstein:2000} Benjamin Grinstein, Detlef R. Nolte and Witold Skiba.
\newblock {\it Adding matter to Poincare invariant branes}.
\newblock {\bf hep-th/0005001}.

\bibitem{Nihei:1999} Takeshi Nihei.
\newblock {\it Inflation in the five-dimensional universe
with an orbifold extra dimension}.
\newblock {\bf hep-ph/9905487}.

\end{thebibliography}
\end{document}